\def\rfr#1{eq. (\ref{#1})}

\def\virg#1{``#1''}

\def\eqi{\begin{equation}}
\def\eqf{\end{equation}}
\def\eqia{\begin{eqnarray}}
\def\eqfa{\end{eqnarray}}
\def\rp#1#2{{#1\over#2}} \def\lb#1{\label{#1}}

\def\bds#1{\boldsymbol{#1}}

\documentclass[11pt,twocolumn]{article}
\usepackage{amsmath,amsthm,amscd,amssymb,wasysym}
\usepackage{latexsym}
\usepackage{graphicx,epsfig}
\usepackage{abstract}
\RequirePackage{color}

\begin{document}

\twocolumn[
	
\noindent{\bf \LARGE{LETSGO: A spacecraft-based mission to accurately measure the solar angular momentum with frame-dragging}}
\\
\\
\\
{L. Iorio$^{\ast}$}\\
{\it $^{\ast}$Ministero dell'Istruzione, dell'Universit\`{a} e della Ricerca (M.I.U.R.)-Istruzione\\ Fellow of the Royal Astronomical Society (F.R.A.S.)\\
Permanent address: Viale Unit$\grave{a}$ di Italia 68
70125 Bari (BA), Italy.  \\ e-mail: lorenzo.iorio@libero.it}

\begin{onecolabstract}
 LETSGO (LEnse-Thirring Sun-Geo Orbiter) is a proposed space-based mission involving the use of a spacecraft moving along a highly eccentric heliocentric orbit perpendicular to the ecliptic. It aims to accurately measure some important physical properties of the Sun and to test some post-Newtonian features of its gravitational field by continuously monitoring the Earth-probe range. Preliminary sensitivity analyses show that, by assuming a cm-level accuracy in ranging to the spacecraft, it would be possible to \textcolor{black}{test}, in principle, the Lense-Thirring effect  at a $\textcolor{black}{\sim 10^{-2}}$ level over a timescale of 2 yr, while the larger Schwarzschild component of the solar gravitational field may be sensed with a relative accuracy of about $10^{-8}-10^{-9}$ during the same temporal interval. The competing range perturbation due to the non-sphericity of the Sun would be a source of systematic error, but it turns out that all the three dynamical features of motion examined affect the Earth-probe range in different ways, allowing for  separating them in real data analyses. The high eccentricity would help in reducing the impact of the non-gravitational perturbations whose disturb would certainly be severe when LETSGO would approach the Sun at just a few solar radii. \textcolor{black}{It can be preliminarily  argued that a drag-free apparatus should perform at a $10^{-8}-10^{-9}\
{\rm m\ s^{-2}\ Hz^{-1/2}}$ level for frequencies of about $10^{-7}$ Hz.} Further studies should be devoted to investigate both the consequences of the non-conservative forces and the actual measurability of the effects of interest by means of extensive numerical data simulations,  parameter estimations and covariance analyses. Also an alternative, fly-by configuration is worth of consideration.
\end{onecolabstract}
Keywords: Classical general relativity; Experimental studies of gravity; Experimental tests of gravitational theories; Main-sequence: intermediate-type stars (A and F); Stellar rotation; Spaceborne and space research instruments, apparatus, and components (satellites, space vehicles, etc.) \\ \\
PACS numbers: 04.20.-q; 04.80.-y; 04.80.Cc; 97.20.Ge; 97.10.Kc; 07.87.+v\\

	]

\section{Introduction}
In this paper, we propose a new spacecraft-based mission, tentatively dubbed LETSGO (LEnse-Thirring Sun-Geo Orbiter). It is mainly, although not exclusively, aimed to  accurately measuring the general relativistic gravitomagnetic field \cite{Thorne1,Rin,MashNOVA} of the rotating Sun through the Lense-Thirring effect \cite{LT} on the \textcolor{black}{highly elliptical} orbital motion of the probe, to be measured by continuous, accurate ranging from the Earth.
\textcolor{black}{See also \cite{Davies} for a suggested space mission for measuring the angular momentum of the Sun based on a different concept. Indeed, Davies \cite{Davies} proposed to use two mutually intercommunicating probes-not necessarily drag-free-moving along identical circular heliocentric orbits, but in opposite directions to exploit the resulting net gravitomagnetic time delay \cite{delay1,delay2}. }
A  concept\footnote{\textcolor{black}{Nonetheless, it also implied the possibility of a direct impact solar probe.}} somewhat analogous to the one presented here was proposed in Ref.~\cite{Rox77} to test other non-gravitomagnetic aspects of the gravitational interaction close to the Sun by ranging with laser and in the X and K bands. For other proposals \textcolor{black}{concerning the possibility of using the Sun's neighborhood to test some aspects of post-Newtonian gravity}, see Refs.~\cite{Israel,Nordt77}.
\textcolor{black}{The proposed 6-months SOREL mission \cite{Israel} was aimed to measuring the Sun's quadrupole mass moment and the general relativistic static component of the solar gravitational field with a drag-free heliocentric probe with a perihelion distance of about $0.28$ AU to be tracked with radio or laser signals from the Earth; the occurrence of Sun occultations (super conjunction) would have afforded to maximize the resulting time delay.}
\textcolor{black}{In Ref.~\cite{Nordt77} Nordtvedt envisaged the use of a stable clock carried onboard a spacecraft in a very eccentric solar orbit with a closest approach of about four solar radii to measure the gravitational red-shift to a part in $10^6$.}
\textcolor{black}{Actually, recent years have seen increasing efforts toward the implementation of the Planetary
Laser Ranging (PLR) technique accurate to about cm level \cite{plr1,plr2,plr3,plr4,plr5,plr6,plr7,plr8,plr9,plr10,plr11}.
}

At present, the existing empirical tests of gravitomagnetism performed in the solar system with either natural or artificial test bodies are few and not conclusive, especially as far as their total accuracy is concerned. For a recent, comprehensive review see, e.g., Ref.~\cite{review}.
\textcolor{black}{The team of the\footnote{\textcolor{black}{See http://einstein.stanford.edu/ on the WEB.}} Gravity Probe B (GP-B) mission \cite{gpb}, implemented in the terrestrial gravitational field with a dedicated orbiting spacecraft, recently reported a direct measurement of another gravitomagnetic effect, i.e. the  Pugh--Schiff  precessions \cite{Pugh59,Schiff60} of four cryogenic gyroscopes carried onboard, in agreement with the predictions of general relativity to within $19\%$ \cite{Eve011}. In 1975 it was proposed by Haas and Ross \cite{Haas75} to measure the same effect with a drag-free spacecraft orbiting the Sun.}
Currently, the major limitation to a direct measurement of the planets' orbital precessions caused by the gravitomagnetic field of the Sun reside in the accuracy with which they can be determined from the planetary observations: indeed, it is nowadays of the same order of magnitude of the relativistic effects themselves \cite{review}.

 In regard to our mission, the direct observable quantity \textcolor{black}{considered here} will be the Earth-probe range $\rho$; \textcolor{black}{in principle, also Doppler range-rate measurements could be considered as well, but we will not do that in the present paper}. We will look at how gravitomagnetism \textcolor{black}{dynamically} affects \textcolor{black}{the range} with respect to  the standard, well tested Newtonian and general relativistic\footnote{Here we refer to the so-called gravitoelectric \cite{MashNOVA}, static component of the gravitational field \cite{Schwa} yielding well known general relativistic phenomena like the geodetic or de Sitter precession  of an orbiting gyroscope \cite{Des}, and the Einstein perihelion precession \cite{Ein}. Several successful empirical checks exist for them since long ago \cite{Will}.}  mechanics due to the orbital motion of LETSGO\footnote{We will not deal with the gravitomagnetic effect on the propagation of the electromagnetic waves between the probe and the terrestrial station(s).}. To this aim, we will, first, numerically integrate the equations of motion in cartesian coordinates of both the Earth and the probe with and without the gravitomagnetic field of the Sun over a suitable time span $\Delta t$. Then, we will compute the time-dependent difference between the ranges computed in both the numerical integrations, i.e. with and without the gravitomagnetic field, in order to obtain a time series for $\Delta\rho_{\rm LT}$ representative of the range shift caused by the solar Lense-Thirring effect \cite{LT}. We will repeat the same analysis also for the non-spherically symmetric component of the Newtonian gravitational field of the Sun due to its quadrupole mass moment $J_2$ \cite{j20,j2} because it is the major source of systematic error of gravitational origin; at present, it is known with a $\sim 10\%$ accuracy \cite{j2acc,j2acc2}.

Basically, the present study is to be intended just as a preliminary concept analysis,  aimed  to set up the scene and to investigate if it is worth pursuing further, more accurate investigations. They should include, for example,  extensive numerical simulations of the probe's data in realistic conditions, and their
processing supplemented by parameter estimation and covariance inspection to effectively test the actual measurability of the Lense-Thirring effect in the proposed scenario. It should check the level of removal of the signal of interest in estimating different sets of solved-for parameters \textcolor{black}{by varying the simulated data sets as well}. Moreover, also the impact of the\footnote{\textcolor{black}{The most important one is the direct solar radiation pressure.}} non-gravitational perturbations of thermal origin, certainly not negligible for an artificial spacecraft moving in a severe environment like the neighborhood of the Sun, \textcolor{black}{and the ways to counteract and/or compensate them \cite{Berto72,Haas75,Roth75,Rox77,Eve78,drag80}} should be accurately investigated in a follow-on of the present study.

Other interplanetary spacecraft-based missions were proposed in the more or recent past to accurately measure the Sun's gravitomagnetic field by means of its direct effects on the propagation of the electromagnetic waves. They are the Laser Astrometric Test of Relativity (LATOR) \cite{lator1}, which aims to directly measure the frame-dragging effect on the light with a $\sim 0.1\%$ accuracy \cite{lator}, and the Astrodynamical Space Test of Relativity Using Optical
Devices I (ASTROD I) \cite{astrod1}, whose goal is to measure the gravitomagnetic component of the time delay with a $10\%$ accuracy \cite{astrod}.

 We also remark that if one assumes the existence of gravitomagnetism as predicted by  general relativity, one can interpret the outcome of our mission as an accurate measurement, in a dynamical and model-independent way, of the angular momentum $\bds S$ of the Sun (see \rfr{bigi} below).
\textcolor{black}{Let us recall that, in the case of a rotating Kerr \cite{Kerr} black hole  of mass $M$, there is a theoretical upper limit \eqi S^{\rm (max)}=\rp{M^2 G}{c},\eqf where $G$ is the Newtonian constant of gravitation and $c$ is the speed of light in vacuum, so that \cite{Shap86}
\eqi S=\chi_g\ S^{\rm (max)},\ \left|\chi_g\right|\leq 1.\eqf If $\left|\chi_g\right| >1$, a naked singularity without a horizon would occur, along with the possibility of causality violations because of closed timelike curves \cite{Chan}. Incidentally, we remark that, although not yet  proven, the cosmic censorship
conjecture \cite{Pen69} states that naked
singularities cannot be formed via the gravitational collapse
of a body. On the contrary, it is known that the dimensionless spin parameter $\chi_g$ of main-sequence stars depends in a non-negligible way on the stellar mass, and it can be well  $\left|\chi_g\right| \gg 1$  \cite{Kraft68,Kraft70,Dicke70,Gray82}; for example, from the analysis in Ref.~\cite{Iorio011}  it can be inferred that $\chi_g\sim 36$ for the star HD15082 (WASP-33) \cite{Gren99}. }
Generally speaking, \textcolor{black}{the angular momentum} can yield relevant information about the inner properties of stars and their activity. Moreover, it
is an important diagnostic for testing theories of stellar formation. The angular momentum can also play a decisive role in stars' evolution, in particular towards the higher mass.
For such topics, see Refs.~\cite{angulo1,angulo2,angulo3,angulo4,angulo5,angulo6}. The asteroseismology technique \cite{seismo1,seismo2,seismo3}, based on the use of all stellar pulsation data, has been used so far to measure the total angular momentum of the Sun and of some other main sequence stars \cite{helio1,dimauro00,antia00,komm03,helio2,yang06,yang08,bi011}. For other determinations, inferred from surface rotation, see Ref.~\cite{livingstone00}.

Finally, we remark that our mission could also be used for accurately, dynamically measuring the solar $J_2$ itself, and  the gravitoelectric part of the general relativistic field of the Sun. In this respect, it could yield greatly improved bounds on the Parameterized Post-Newtonian (PPN) parameters $\beta$ and $\gamma$ \cite{ppn} compared  to the present-day ones \cite{Will}.
%
\section{The dynamical accelerations}
\textcolor{black}{The exterior  spacetime metric of the Sun can  adequately be described within the PPN approximation \cite{Sof89}.}
\textcolor{black}{
Within such a framework,}
 the \textcolor{black}{general relativistic} stationary gravitomagnetic field $\bds B_g$ of a slowly rotating body with proper angular momentum $\bds S$ is, at great distance $r$ from it, \cite{Thorne1,Thorne,Mash01}
\eqi\bds B_g=-\rp{G}{cr^3}\left[\bds S-3\left(\bds S\bds\cdot\bds{\hat{r}}\right)\bds{\hat{r}}\right],\lb{bigi}\eqf
\textcolor{black}{where $\bds{\hat{r}}$ is the unit position vector directed from the body to a point in space outside it, and the central dot $\bds\cdot$ denotes the usual scalar product between  three dimensional vectors.}
Notice that \rfr{bigi}\textcolor{black}{, which can also be thought as arising from a weak-field, linearized version of the Kerr metric,} exhibits an azimuthal symmetry, i.e. it is the same in all planes containing $\bds S$. The gravitomagnetic field  of \rfr{bigi} affects a test particle moving with velocity $\bds v$ \textcolor{black}{with respect to the rotating body through} a non-central, Lorentz-like acceleration\footnote{\textcolor{black}{The general relativistic multiplicative factor 2 in \rfr{alt} corresponds to $1+\gamma$ in PPN formalism.}} \cite{IERS}
\eqi\bds A_{\rm LT}=-\textcolor{black}{2}\left(\rp{\bds v}{c}\right)\bds\times \bds B_g,\lb{alt}\eqf which is analogous to the one felt by a moving electric charge in a magnetic field in the framework of the Maxwellian electromagnetism\textcolor{black}{; here the central cross $\bds\times$ denotes the usual vector product between three dimensional vectors}.
Helioseismology yields\footnote{\textcolor{black}{Incidentally, it implies for the Sun's dimensionless spin parameter $\chi_g=0.2$.}}, on average, \cite{helio1,dimauro00,antia00,komm03,helio2,yang06,yang08,bi011}\eqi S = 1.92\times 10^{41}\ {\rm kg\ m^2\ s^{-1}}\lb{seismol}\eqf for the Sun, with an uncertainty at a percent level or less, so that \rfr{alt} can be viewed as a tiny perturbation of the usual Newtonian monopole \textcolor{black}{acceleration}
\eqi\bds A_{\rm N}=-\rp{GM}{r^2}\bds{\hat{r}},\eqf where $M$ is the mass of a solar-type main-sequence star ($GM_{\odot}=1.327\times 10^{20}\ {\rm m^3\ s^{-2}}$ \cite{Allen}).
%
%
%
%
%
%
%

\textcolor{black}{
Concerning the Schwarzschild-type gravitoelectric, static component of the solar gravitational field, the \textcolor{black}{test particle acceleration induced by} it can be modeled as \cite{Sof89,IERS}
\eqi \bds{A}_{\rm 1PN}=\rp{GM}{c^2r^3}\left[\left(\rp{4GM}{r}-v^2\right)\bds r+4\left(\bds r\bds\cdot\bds v\right)\bds v\right]\lb{agel}\eqf
at 1PN level in general relativity\footnote{\textcolor{black}{In the PPN formalism the terms $GM/r$, $v^2$ and $\left(\bds r\bds\cdot\bds v\right)$ in \rfr{agel} are multiplied by $2\left(\beta+\gamma\right)$, $\gamma$ and $2\left(1+\gamma\right)$, respectively.}}, \textcolor{black}{where $\bds{r}$ is the position vector of the test particle with respect to the Sun}. \textcolor{black}{Its magnitude} is larger than
%
%
%
%
$A_{\rm LT}$, but, as already stated, several predictions of \rfr{agel}
were successfully tested with a variety of techniques: as a consequence, $\gamma$ and $\beta$ are nowadays known at a $10^{-5}-10^{-4}$ level \cite{Will}.
}

In principle,  a 2PN term of order $\mathcal{O}(c^{-4})$ exists as well \cite{2pn1,2pn2,2pn3,2pn4}. However, we will not consider it since its effects are completely negligible. Suffice it to say that the general relativistic 2PN secular precession of the perihelion $\omega$ of a test particle orbiting $M$ is \cite{2pn3}
\eqi\dot\omega_{\rm 2PN}\sim \rp{3(GM)^2 n}{c^4 a^2(1-e^2)^2},\lb{2pnacc}\eqf where $a$ is the semi-major axis, $e$ is the eccentricity and $n\doteq\sqrt{GM/a^3}$ is the
Keplerian mean motion.
%
%
%
%
%

The first even zonal harmonic coefficient $J_2$ \cite{j20,j2} of the multipolar expansion of the non-spherically symmetric part of the Newtonian gravitational potential of the Sun accounting for its oblateness \textcolor{black}{according to}
\eqi \textcolor{black}{U_{\rm obl} = -\rp{GM}{r}J_2\left(\rp{R}{r}\right)^2 P_2\left(\bds{k}\bds\cdot\bds{\hat{r}}\right),} \eqf \textcolor{black}{where $P_2$ is the Legendre polynomial of degree 2, $R$ is the solar equatorial radius, and $\bds k$ is a unit vector along the Sun's rotational axis}, yields the following \textcolor{black}{classical} perturbing acceleration\footnote{\textcolor{black}{The nabla operator $\bds\nabla$ is a symbolic vector whose components are the partial derivatives $\partial/\partial x_i,\ i=1,2,3$ with respect to the spatial coordinates $x_i, i=1,2,3$.}} \cite{Vrbik}
\begin{align} \bds A_{\rm obl} \lb{AJ2}\nonumber & = \textcolor{black}{\bds{\nabla} U_{\rm obl}} = \\ \nonumber \\
&- \rp{3J_2 R^2 GM}{2 r^4}\left\{\left[1-5\left(\bds{\hat{r}}\bds\cdot\bds k\right)^2\right]\bds{\hat{r}}+2\left(\bds{\hat{r}}\bds\cdot\bds k\right)\bds k\right\}.\end{align}
For the Sun it is \cite{Allen,j2acc2}
\begin{equation}
\left\{
\begin{array}{lll}
J_2 & = & 2\times 10^{-7}, \\ \\
R & = & 6.96\times 10^8\ {\rm m}=0.00465\ {\rm au},\\ \\
\bds k & = & \{0.12,-0.03,0.99\}.
\end{array}
\right.\lb{compos}
\end{equation}
Concerning $\bds k$, we adopted a Sun-centered frame $K$ with the mean ecliptic and equinox at J2000.0 epoch as reference $\{xy\}$ plane and $x$ direction, respectively; the values of its components in \rfr{compos} come from the fact that the right ascension $\alpha_0$ and declination $\delta_0$ of the Sun's north pole of rotation with respect to the mean terrestrial equator at J2000.0 are $\alpha_0=286.13\ {\rm deg},\ \delta_0=63.87$ deg \cite{aldec}, respectively, and the obliquity of the Earth's equator to the ecliptic at J2000.0 is $\varepsilon=23.439$ deg \cite{fuku}. \textcolor{black}{The vectors $\bds r,\bds v,$ and their scalar product $\bds r\bds\cdot\bds v$ entering \rfr{agel} and \rfr{AJ2} refers just to $K$.}
The impact of the other Sun's even zonals of higher degrees is negligible. At present, there are evaluations only for the second even zonal harmonic coefficient $J_4$, whose magnitude should be of the order of $10^{-7}$ as well \cite{j4}: its dynamical effect is completely negligible with respect to $A_{J_2}$ because of an additional multiplicative factor $(R/r)^2$.

\textcolor{black}{
In principle, there is also a variety of general relativistic orbital effects of order $\mathcal{O}(c^{-2})$ pertaining $J_2$ \cite{relobl1,relobl2}. However, they are all negligible since they are of the order of \cite{relobl1,relobl2}
\eqi A_{\rm rel\ obl}\sim \left(\rp{GM}{c^2 a}\right)A_{\rm obl}.\eqf
}

The uncertainties in physical parameters of the Sun like its gravitational parameter $\mu\doteq GM$ and its equatorial radius $R$ are of no concern since the resulting mismodeling in the general relativistic signals of interest is negligible. Indeed, it is \cite{j2acc2}
\eqi \sigma_\mu=1\times 10^9\ {\rm m^3\ s^{-2} }\eqf corresponding to a fractional uncertainty in $\mu$
as small as \eqi \rp{\sigma_{\mu}}{\mu}=7\times 10^{-12}.\lb{fracMU}\eqf
Concerning $R$, it turns out that the fractional uncertainty in it can be evaluated to be of the order of \cite{radius1,radius2,radius3}
\eqi \rp{\sigma_R}{R}\sim 10^{-4}-10^{-5}.\lb{fracR}\eqf
An application of the figures in \rfr{fracMU} and \rfr{fracR} to those listed in Table \ref{resume1}, Table \ref{resume2}, and Table \ref{resume3} clearly shows how the resulting biased signatures are negligible with respect to the Lense-Thirring ones.

\textcolor{black}{
As a final remark, it is important to stress that the relative accuracy of about $10^{-2}$  with which $S_{\odot}$ is presently known from helioseismology implies that the accuracy of a LETSGO-based test of the Lense-Thirring effect could not be better than such a figure, even if the mission design and the technological capabilities would allow, in principle, to obtain a better  measurement precision.
}
\section{Numerical analysis}
Let us assume that a probe is launched from the Earth with a geocentric velocity $\bds v^{'}_{\rm p}$ with a magnitude almost equal to  $v_{\oplus}$, in such a way that its heliocentric velocity $\bds v_{\rm p}$  has a magnitude smaller than the terrestrial one. Here we neglect practical considerations concerning how to actually implement such an orbital insertion: it could also be assumed that the probe is launched with a different velocity, and it is finally brought to such a configuration by means of subsequent orbital maneuvers and/or one or more flybys with other planets. Anyway, if at a given epoch  $v_{\rm p}<v_{\oplus}$, then the probe will follow a more eccentric elliptic path with respect to the terrestrial one which will bring it very close to the Sun, depending on $v_{\rm p}$.
Concerning the direction of $\bds v_{\rm p}$, it must not necessarily coincide with that of $\bds{v}_{\oplus}$: it can be suitably chosen in order to enhance the relativistic effects of interest with respect to the classical ones acting as disturbing biases.

In Figure \ref{figura1} we depict  the nominal Schwarzschild and Lense-Thirring range signals $\Delta\rho_{\rm Sch},\ \Delta\rho_{\rm LT}$, and the mismodeled $J_2$ dynamical range perturbation $\Delta\rho_{J_2}$ for a probe's elliptical path with $a=0.51$ au and $e=0.92$. The integration time chosen is $\Delta t=2$ yr. We assumed a launch height of 200 km with respect to the Earth's surface.
\begin{figure*}[ht!]
\centering
\begin{tabular}{c}
\epsfig{file=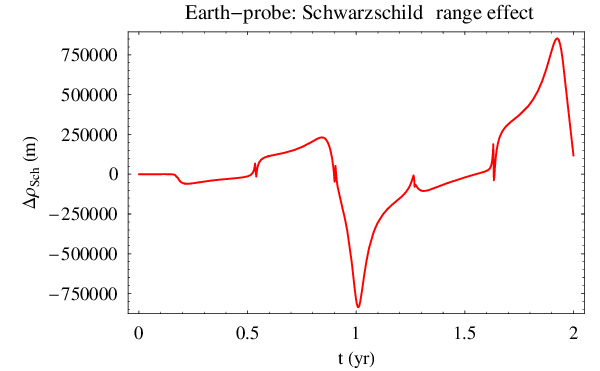,width=0.35\linewidth,clip=} \\
\epsfig{file=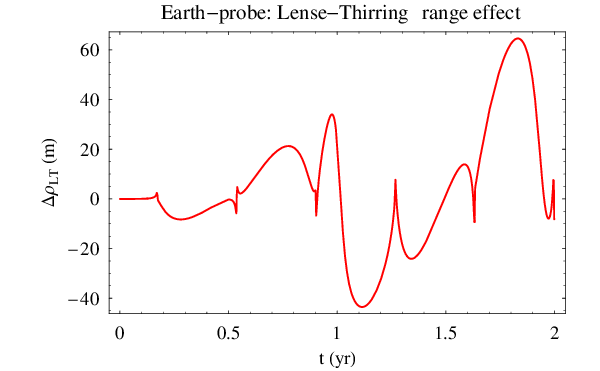,width=0.35\linewidth,clip=} \\
\epsfig{file=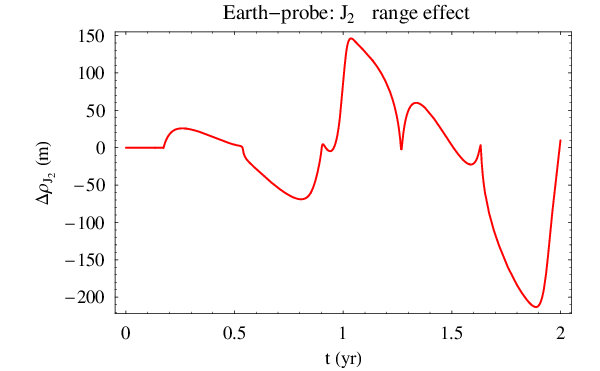,width=0.35\linewidth,clip=}\\
\end{tabular}
\caption{Differences $\Delta\rho_{\rm Sch}$, $\Delta\rho_{\rm LT}$, and $\Delta\rho_{J_2}$, in m, of the numerically integrated  Earth-probe ranges  with and without the nominal general relativistic Schwarzschild and Lense-Thirring perturbations (top and middle panels), and the classical  dynamical perturbation due to the mismodeled even zonal harmonic $J_2$ (bottom panel). A $10\%$ mismodeling in $J_2$ was adopted. The initial conditions are common to both the perturbed and un-perturbed integrations. For the Earth they were retrieved from the WEB interface HORIZONS by NASA JPL at epoch J2000.0. The initial state vector of the probe p is $x^{\rm p}_0=\left(1+d/r_0^{\oplus}\right)x_0^{\oplus},y^{\rm p}_0=\left(1+d/r_0^{\oplus}\right)y_0^{\oplus},z^{\rm p}_0=\left(1+d/r_0^{\oplus}\right)z_0^{\oplus},\dot x^{\rm p}_0=0,\dot y^{\rm p}_0=0,\dot z^{\rm p}_0=0.28 v_0^{\oplus}$; it corresponds to $a^{\rm p}_0=0.51$ au, $e_0^{\rm p}=0.920$, $I_0^{\rm p}=90$ deg. We used $d=R_{\oplus}+h$, with the launch height given by $h=200$ km. The time span is $\Delta t=2$ yr. The minimum distance of the probe from the Sun turns out to be $r^{\rm p}_{\rm min}=0.0385\ {\rm au}=8.29R$.}\lb{figura1}
\end{figure*}
\begin{table*}[ht!]
\caption{Peak-to-peak maximum amplitude $|\Delta\rho|^{\rm max}$, mean $\left\langle\Delta\rho\right\rangle$ and variance $\sigma_{\Delta\rho}$, in m, of the range signals of Figure \ref{figura1} caused by the general relativistic nominal effects (Schwarzschild and Lense-Thirring), and by the mismodeled classical perturbation due to the Sun's first even zonal coefficient $J_2$, assumed to be known at a $10\%$ level. The integration interval is $\Delta t=2$ yr. The minimum distance reached by the probe is $8.29$ solar radii. }
\label{resume1}
\centering
\bigskip
\begin{tabular}{llll}
\hline\noalign{\smallskip}
Dynamical orbital effect & $|\Delta\rho|^{\rm max}$ (m) & $\left\langle\Delta\rho\right\rangle$ (m)&  $\sigma_{\Delta\rho}$ (m)  \\
\noalign{\smallskip}\hline\noalign{\smallskip}
Schwarzschild & $1,688,683.4$ & $40,057.7$ & $282,599.0$ \\
Lense-Thirring & $108.1$ & $4.5$ & $24.5$\\
$J_2$ ($\delta J_2/J_2=0.1$) & $359.2$ & $-13.5$ & $81.5$\\
\noalign{\smallskip}\hline\noalign{\smallskip}
\end{tabular}
\end{table*}
 Table \ref{resume1} summarizes the main quantitative features of the three signatures investigated. By assuming a cm-level accuracy in ranging to LETSGO, the \textcolor{black}{nominal} gravitomagnetic effect would be, in principle, detectable with an accuracy of about one part per\footnote{\textcolor{black}{The actual accuracy of a Lense-Thirring test would be set by \rfr{seismol} affecting its theoretical prediction.}} $1,000-10,000$, while the larger Schwarzschild shift would be detectable at a $10^{-8}-10^{-9}$ level. It is important to notice that the temporal patterns of the three time series are different, so that they could be more easily separated \textcolor{black}{with standard filtering techniques (Kalman filtering, etc.)}, especially as far as the Lense-Thirring and the $J_2$ signals are concerned. It is important since they are about of the same order of magnitude, being the mismodeled classical effect $3-4$ times larger than the gravitomagnetic one.
From the practical point of view, it is important to notice that the minimum distance from the Sun is $8.29$ solar radii. It is similar to the one of the NASA mission\footnote{See http://solarprobe.jhuapl.edu/ on the WEB. It aims to understand how the solar  corona is heated and how the solar wind is accelerated.} Solar Probe Plus \cite{SP1}, recently approved and scheduled for a launch in 2018, which should reach a closest distance of $8.5$ solar radii, thus facing a temperature of $2000^{\circ}$ Celsius.
\textcolor{black}{We remark that LETSGO would certainly benefit from the studies conducted for such a mission, especially as far as the non-gravitational forces are concerned \cite{SP2}.}
\textcolor{black}{As a preliminary remark, we notice} that the high value of the eccentricity of LETSGO would be helpful in greatly reducing the impact of the non-gravitational perturbations \cite{Milani}.
At the moment, the spacecraft which came closest to the Sun so far is the Helios-2 spacecraft \cite{helios2}, which reached a minimum distance of $0.29\ {\rm au} =63.2R$  in 1976.
\textcolor{black}{As a naive, order-of-magnitude evaluation, it may be argued that a hypothetical drag-free device for LETSGO should be able to counteract non-gravitational accelerations down to a\footnote{\textcolor{black}{The orbital period of LETSGO would be $P_{\rm orb}=133$ d, corresponding to a frequency  $\nu_{\rm orb}=9\times 10^{-8}$ Hz. The maximum value of the LETSGO's nominal Lense-Thirring acceleration, experienced  at perihelion, would be as large as $A_{\rm LT}\sim 2\times 10^{-10}$  m s$^{-2}$.}} $\sim 10^{-8}-10^{-9}\ {\rm m\ s^{-2}\ Hz^{-1/2}}$ level for orbital frequencies of the order of $\sim 10^{-7}$ Hz in order to achieve a $\sim 10^{-3}$ measurement. As a comparison, the performance of the LISA drag-free apparatus should be less than \cite{dragfree,ashby} $10^{-15}\ {\rm m\ s^{-2}\ Hz^{-1/2}}$ down to $10^{-4}$ Hz. } \textcolor{black}{As a rough order-of-estimate of the magnitude of the perturbing acceleration due to the Solar radiation pressure, it can be posed $A_{\rm SRP}\sim 10^{17}\ {\rm kg\ m\ s^{-2}} r^{-2}\left(\Sigma/m\right)$, where $10^{17}\ {\rm kg\ m\ s^{-2}}$ is the approximate value of the Solar radiation constant, and $\Sigma/m$ is the area-to-mass ratio of LETSGO. For $r^{\rm p}_{\rm min}=4.31R$ (Figure \ref{figura2}, Table \ref{resume2}), it yields $A_{\rm SRP}^{\rm max}\sim 1\times 10^{-2}\ {\rm kg\ m^{-1}\ s^{-2}}\left(\Sigma/m\right)$; for $r^{\rm p}_{\rm min}=12.67R$ (Figure \ref{figura3}, Table \ref{resume3}), it is $A_{\rm SRP}^{\rm max}\sim 1\times 10^{-3}\ {\rm kg\ m^{-1}\ s^{-2}}\left(\Sigma/m\right)$.}

The dependence on the initial velocity $\bds v_0^{\rm p}$ is particularly relevant, and it  can, in principle, be suitably tuned in order to improve the practical feasibility of the mission, especially as far as the severe conditions encountered at so close distances from the Sun are concerned. For example, it can be shown that by choosing $\bds v_0^{\rm p}=0.2 v_0^{\oplus}\bds{\hat{z}}$ it is possible to obtain a smaller minimum distance of $4.3$ solar radii
by enhancing the magnitude of the range signatures as depicted in Figure \ref{figura2}; see Table \ref{resume2} for the quantitative features of the competing signals.
\begin{figure*}[ht!]
\centering
\begin{tabular}{c}
\epsfig{file=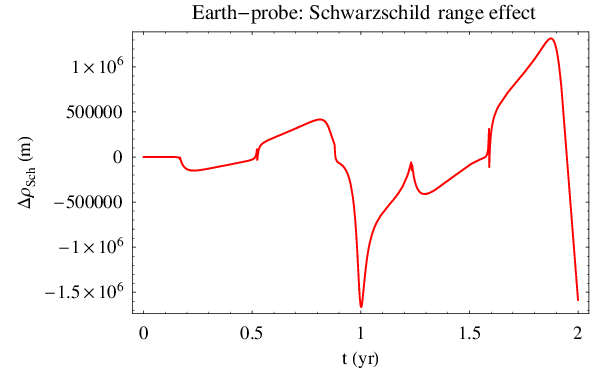,width=0.35\linewidth,clip=} \\
\epsfig{file=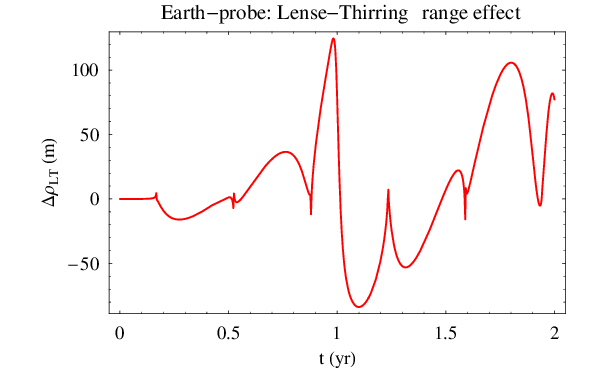,width=0.35\linewidth,clip=} \\
\epsfig{file=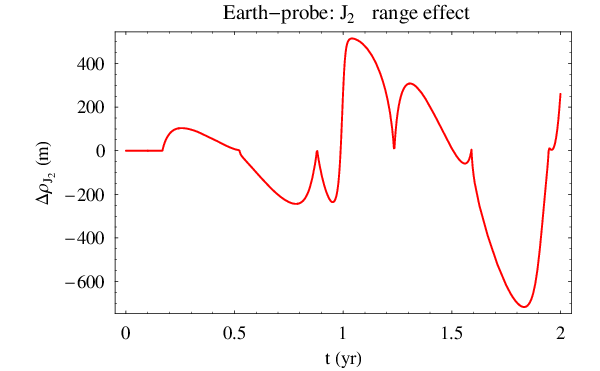,width=0.35\linewidth,clip=}\\
\end{tabular}
\caption{Differences $\Delta\rho_{\rm Sch}$, $\Delta\rho_{\rm LT}$, and $\Delta\rho_{J_2}$, in m, of the numerically integrated  Earth-probe ranges  with and without the nominal general relativistic Schwarzschild and Lense-Thirring perturbations (top and middle panels), and the classical  dynamical perturbation due to the mismodeled even zonal harmonic $J_2$ (bottom panel). A $10\%$ mismodeling in $J_2$ was adopted. The initial conditions are common to both the perturbed and un-perturbed integrations. For the Earth they were retrieved from the WEB interface HORIZONS by NASA JPL at epoch J2000.0. The initial state vector of the probe p is $x^{\rm p}_0=\left(1+d/r_0^{\oplus}\right)x_0^{\oplus},y^{\rm p}_0=\left(1+d/r_0^{\oplus}\right)y_0^{\oplus},z^{\rm p}_0=\left(1+d/r_0^{\oplus}\right)z_0^{\oplus},\dot x^{\rm p}_0=0,\dot y^{\rm p}_0=0,\dot z^{\rm p}_0=0.2 v_0^{\oplus}$; it corresponds to $a^{\rm p}_0=0.50$ au, $e_0^{\rm p}=0.959$, $I_0^{\rm p}=90$ deg. We used $d=R_{\oplus}+h$, with the launch height given by $h=200$ km. The time span is $\Delta t=2$ yr. The minimum distance of the probe from the Sun turns out to be $r^{\rm p}_{\rm min}=0.020\ {\rm au}=4.31R$.}\lb{figura2}
\end{figure*}
\begin{table*}[ht!]
\caption{Peak-to-peak maximum amplitude $|\Delta\rho|^{\rm max}$, mean $\left\langle\Delta\rho\right\rangle$ and variance $\sigma_{\Delta\rho}$, in m, of the range signals of Figure \ref{figura2} caused by the general relativistic nominal effects (Schwarzschild and Lense-Thirring), and by the mismodeled classical perturbation due to the Sun's first even zonal  $J_2$, assumed to be known at a $10\%$ level. The integration interval is $\Delta t=2$ yr. The minimum distance reached by the probe is $4.31$ solar radii. }
\label{resume2}
\centering
\bigskip
\begin{tabular}{llll}
\hline\noalign{\smallskip}
Dynamical orbital effect & $|\Delta\rho|^{\rm max}$ (m) & $\left\langle\Delta\rho\right\rangle$ (m)&  $\sigma_{\Delta\rho}$ (m)  \\
\noalign{\smallskip}\hline\noalign{\smallskip}
Schwarzschild & $2,980,019.3$ & $30,744.1$ & $541,472.0$ \\
Lense-Thirring & $208.3$ & $9.4$ & $4.0$\\
$J_2$ ($\delta J_2/J_2=0.1$) & $1,231.3$ & $-38.7$ & $288.4$\\
\noalign{\smallskip}\hline\noalign{\smallskip}
\end{tabular}
\end{table*}
Apart from the certainly much more severe challenges posed by the extreme closeness to the Sun's photosphere, the scenario of Table \ref{resume2} is slightly less favorable than the one in Table \ref{resume1}, especially as far as the Lense-Thirring effect is concerned. Indeed, the mismodeled $J_2$ signal would be larger; moreover, the increase of the gravitomagnetic range shift in terms of its potential measurability would be rather modest with respect to the safer scenario of Table \ref{resume1}. Also the Schwarzschild-to-$J_2$ ratio would be less favorable.
On the other hand, $\bds v_0^{\rm p}=0.35 v_0^{\oplus}\bds{\hat{z}}$ yields a minimum distance of $12.7$ solar radii, with the range signals of interest depicted in Figure \ref{figura3}: their quantitative features are summarized in Table \ref{resume3}.
\begin{figure*}[ht!]
\centering
\begin{tabular}{c}
\epsfig{file=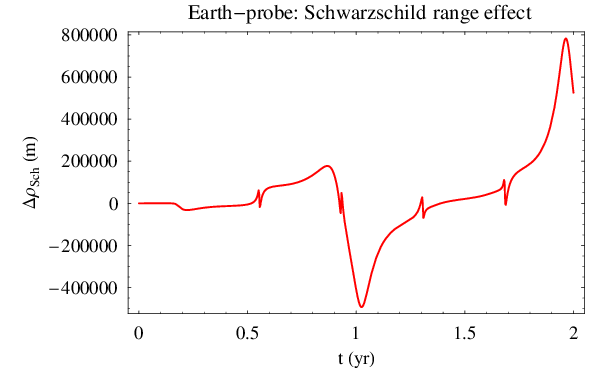,width=0.35\linewidth,clip=} \\
\epsfig{file=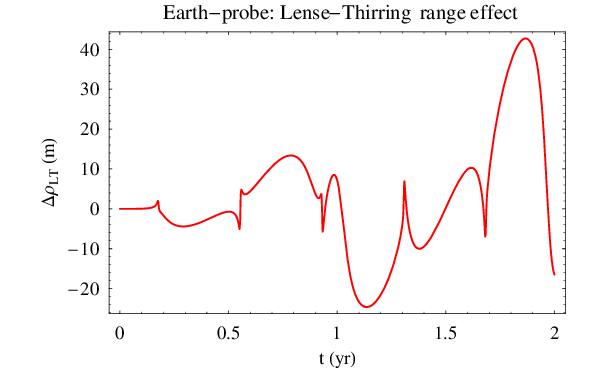,width=0.35\linewidth,clip=} \\
\epsfig{file=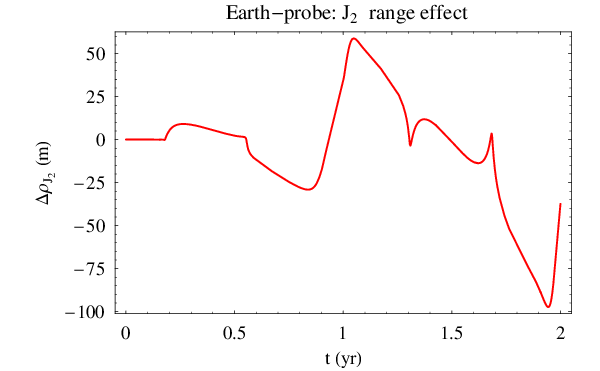,width=0.35\linewidth,clip=}\\
\end{tabular}
\caption{Differences $\Delta\rho_{\rm Sch}$, $\Delta\rho_{\rm LT}$, and $\Delta\rho_{J_2}$, in m, of the numerically integrated  Earth-probe ranges  with and without the nominal general relativistic Schwarzschild and Lense-Thirring perturbations (top and middle panels), and the classical  dynamical perturbation due to the mismodeled even zonal harmonic $J_2$ (bottom panel). A $10\%$ mismodeling in $J_2$ was adopted. The initial conditions are common to both the perturbed and un-perturbed integrations. For the Earth they were retrieved from the WEB interface HORIZONS by NASA JPL at epoch J2000.0. The initial state vector of the probe p is $x^{\rm p}_0=\left(1+d/r_0^{\oplus}\right)x_0^{\oplus},y^{\rm p}_0=\left(1+d/r_0^{\oplus}\right)y_0^{\oplus},z^{\rm p}_0=\left(1+d/r_0^{\oplus}\right)z_0^{\oplus},\dot x^{\rm p}_0=0,\dot y^{\rm p}_0=0,\dot z^{\rm p}_0=0.35 v_0^{\oplus}$; it corresponds to $a^{\rm p}_0=0.52$ au, $e_0^{\rm p}=0.875$, $I_0^{\rm p}=90$ deg. We used $d=R_{\oplus}+h$, with the launch height given by $h=200$ km. The time span is $\Delta t=2$ yr. The minimum distance of the probe from the Sun turns out to be $r^{\rm p}_{\rm min}=0.058\ {\rm au}=12.67R$.}\lb{figura3}
\end{figure*}
\begin{table*}[ht!]
\caption{Peak-to-peak maximum amplitude $|\Delta\rho|^{\rm max}$, mean $\left\langle\Delta\rho\right\rangle$ and variance $\sigma_{\Delta\rho}$, in m, of the range signals of Figure \ref{figura3} caused by the general relativistic nominal effects (Schwarzschild and Lense-Thirring), and by the mismodeled classical perturbation due to  the Sun's first even zonal $J_2$, assumed to be known at a $10\%$ level. The integration interval is $\Delta t=2$ yr. The minimum distance reached by the probe is $12.67$ solar radii. }
\label{resume3}
\centering
\bigskip
\begin{tabular}{llll}
\hline\noalign{\smallskip}
Dynamical orbital effect & $|\Delta\rho|^{\rm max}$ (m) & $\left\langle\Delta\rho\right\rangle$ (m)&  $\sigma_{\Delta\rho}$ (m)  \\
\noalign{\smallskip}\hline\noalign{\smallskip}
Schwarzschild & $1,275,165.4$ & $35,013.4$ & $195,204.0$ \\
Lense-Thirring & $67.3$ & $2.9$ & $14.9$\\
$J_2$ ($\delta J_2/J_2=0.1$) & $156.1$ & $-6.2$ & $33.1$\\
\noalign{\smallskip}\hline\noalign{\smallskip}
\end{tabular}
\end{table*}
 It shows a better scenario than Table \ref{resume1} and, especially, Table \ref{resume2}. Indeed, given a cm-level accuracy in measuring the range of the probe, the sensibility to the Lense-Thirring signal experiences just a small degradation with respect to Table \ref{resume1}, remaining at the level of $10^{-3}-10^{-4}$. On the other hand, the aliasing due to the $10\%$ mismodeling in the $J_2$ signal is about  twice the gravitomagnetic one, i.e. approximately $1.4-1.8$ times smaller than in Table \ref{resume1}. The situation for the Schwarzschild range perturbation is improved as well. Indeed, while, on the one hand, its measurability would remain at about $10^{-8}-10^{-9}$, on the other hand, it would be larger than the mismodeled $J_2$ effect by a factor $2$ with respect to Table \ref{resume1}.

It can be shown that the trajectories lying in the ecliptic plane are less convenient; for this reason we do not depict figures dealing with such scenarios.

Given that the ranging accuracy actually obtainable depends on how well the orbit of the Earth is known, a consideration is in order as far as the choice of the time interval of the analysis is concerned. It cannot be too long since, otherwise, the biasing action of the asteroidal belt on the Earth would have a non-negligible impact on its orbital motion \cite{aster}.
\section{Summary and Conclusions}
We proposed a new space-based mission, named LETSGO, aimed to accurately measure some key physical properties of the Sun by  continuously monitoring  the distance between the Earth and  a spacecraft moving along a highly eccentric heliocentric orbit. Its data could also be used to accurately test some post-Newtonian features of the solar gravitational field, like its gravitomagnetic component, through their impact on the orbital motion of the probe.

We just performed  a preliminary sensitivity analysis dealing with only the main gravitational effects on the probe's dynamics; actually, we did not check the effective measurability of the investigated features of motion by simulating data points and fitting dynamical models to them. We  took into account neither the impact of the non-gravitational perturbations nor of the orbital maneuvers. The effects of the aforementioned post-Newtonian terms on the propagation of the electromagnetic waves linking the Earth and LETSGO were neglected as well. Further, dedicated analyses may treat them in detail.

It turned out that, by assuming an overall cm-level accuracy in determining the probe's orbit, the \textcolor{black}{nominal} Lense-Thirring effect on it would be measurable, in principle, at a $10^{-3}-10^{-4}$ level, \textcolor{black}{ although the present-day accuracy with which the Sun's angular momentum is known  sets a  $\sim 10^{-2}$ limit to a test of the Lense-Thirring effect}. The larger gravitoelectric, Schwarzschild-type part of the Sun's field may be detected at about $10^{-8}-10^{-9}$ level.
The Earth-LETSGO range would be affected by the Newtonian non-spherically symmetric component of the solar field in such a way that it could be accurately measured as well. We showed that these three competing dynamical orbital effects have the important property that their temporal patterns are quite different, thus facilitating their separation in data processing and likely a more accurate determination. The choice of the initial heliocentric velocity of the probe would be important;  orbits lying in planes perpendicular to the ecliptic would be favored with respect to ecliptical trajectories.  The high values of the eccentricity, which allows for distances of closest approach to the Sun of just some solar radii, would allow to greatly reduce the averaged effects of the non-gravitational perturbations. \textcolor{black}{They should be counteracted by a hypothetical drag-free apparatus down to a $10^{-8}-10^{-9}\
{\rm m\ s^{-2}\ Hz^{-1/2}}$ level for frequencies of about $10^{-7}$ Hz}. A time span of just a few years would be needed in order to prevent the long-term corrupting impact of the asteroids on the Earth's orbital motion.

In conclusion, the scenario outlined for LETSGO is promising, deserving further investigations. For example, also an alternative fly-by configuration \textcolor{black}{and Doppler range-rate measurements} would be worth of consideration.


\end{document}